\documentclass[letterpaper, 10 pt, conference]{ieeeconf}           

\IEEEoverridecommandlockouts                              

\usepackage{blindtext}
\usepackage{enumerate}
\usepackage{verbatim}
\usepackage{fancyhdr}
\usepackage{float}
\usepackage{mathrsfs}
\usepackage{listings}
\usepackage{graphics} 
\usepackage{epsfig} 
\usepackage{mathptmx} 
\usepackage{times} 
\usepackage{amsmath} 
\usepackage{amssymb}  
\usepackage{amsfonts}
\usepackage{cite}
\usepackage{epstopdf}
\usepackage{tcolorbox}
\usepackage{graphicx,setspace}
\usepackage{bbm}
\usepackage{comment}
\usepackage{subcaption}
\usepackage{algorithm, algpseudocode}
\usepackage[colorinlistoftodos]{todonotes}

\title{\LARGE \bf
	Rapid Circadian Entrainment in Models of Circadian Genes Regulation}

\author{Jiawei Yin, Agung Julius and John T. Wen}
\begin{document}
	
	\maketitle
	\thispagestyle{empty}
	\pagestyle{empty}
	
	\begin{abstract}
		
		The light-based minimum-time circadian entrainment problem for mammals, Neurospora, and Drosophila is studied based on the mathematical models of their circadian gene regulation. These models contain high order nonlinear differential equations. Two model simplification methods are applied to these high-order models: the phase response curves (PRC) and the Principal Orthogonal Decomposition (POD). The variational calculus and a gradient descent algorithm are applied for solving the optimal light input in the high-order models. As the results of the gradient descent algorithm rely heavily on the initial guesses, we use the optimal control of the PRC and the simplified model to initialize the gradient descent algorithm. In this paper, we present: (1) the application of PRC and direct shooting algorithm on high-order nonlinear models; (2) a general process for solving the minimum-time optimal control problem on high-order models; (3) the impacts of minimum-time optimal light on circadian gene transcription and protein synthesis.
		
	\end{abstract}
	
	\section{Introduction}
	
	
	{\let\thefootnote\relax\footnote{{The authors are with the Light Enabled Systems and Application (LESA) Engineering Research Center, Rensselaer Polytechnic Institute, Troy, NY, USA. Research reported in this paper is supported by the NSF through grant number EEC-0812056 and the ARO through
	grant number W911NF-13-1-0265. The authors would like to thank Dr. Wei Qiao for his contribution to the initial portion of this research as part of his postdoctoral research. Email:\texttt{ \{yinj4,juliua2,wenj\}@rpi.edu.}}}}A wide range of biological processes in terrestrial species display endogenous oscillations with a period of about 24 hours. We call these processes the circadian rhythm. The circadian rhythm is in close connection with the numerous biological processes, including the sleep-wake cycle, blood pressure, hormone secretion, circadian gene transcription and protein synthesis\cite{Crosthwaite1997}\cite{Goldbeter1996}.
	
	Maintenance of the circadian rhythm is essential for mammals, insects, and fungi. Being the most powerful influence on the circadian system, light has been used in circadian entrainment in the previous works\cite{Jewett1991}\cite{Bagheri2008}. In view of control theory, the circadian entrainment problem is usually treated as a tracking problem, in which light is used to drive the subject's circadian state to a reference trajectory\cite{Zhang2013}\cite{Julius2017}. The light input for circadian entrainment is usually proposed based on mathematical models of the circadian system. One of the most widely-used circadian models is proposed by Kronauer et al.\cite{Kronauer1999}\cite{Jewett1999}. In this model, the dynamics of the core body temperature is modeled as an oscillator to represent the circadian clock. Light strategies for rapid circadian entrainment on the Kronauer model were given in previous works \cite{Serkh2014}, \cite{Julius2018}, \cite{Zhang2012} \cite{Zhang2016}.
	
	The molecular mechanism of the circadian system has also been studied and the circadian genes expression activities are proposed to represent the circadian oscillator. Leloup et al. \cite{Goldbeter98}\cite{Leloup99} presented a 10th-order ODE Drosophila circadian model, which contains the dynamics of the Period (PER) and Timeless (TIM) proteins. A molecular model of Neurospora was given by Leloup and Gonze \cite{Leloup99} in terms of the $frq$ mRNA transcription and FRQ protein synthesis. Numerous models of mammalian circadian rhythm were formulated based on the negative feedback loop in the $Period\ (Per)$ and $Cryptochrome\ (Cry)$ genes transcription\cite{Sabine2004}\cite{Leloup2003}.
	
	In this paper, we seek to solve the minimum-time entrainment problem, i.e., how to use light to synchronize the phase of the circadian rhythm with a reference phase as quickly as possible. We have recently solved this problem for the empirical Kronauer circadian model, which has 3 states (2 oscillatory states and 1 receptor state) \cite{Julius2018}. Our approach, based on a functional gradient descent algorithm, searches for the optimal light input for the full-order (3rd order) model using the optimal light inputs of some reduced-order models as \emph{initial guesses}. We defined two reduced order models: (1) A first-order model that captures only the phase dynamics of the oscillator, and (2) A second-order model that ignores the receptor state dynamics. The first-order model is accurate around the limit cycle of the system, while the second-order model can capture the amplitude dynamics of the oscillation in addition to the phase dynamics. We demonstrated that these initial guesses led to the optimal solution for the full-order system. The circadian models of mammals and Drosophila that are based on circadian genes regulation dynamics have higher orders (7 or more states). For these systems, the optimal light strategy for minimum-time entrainment is hard to calculate directly. Extending the ideas from \cite{Julius2018}, we propose a general solution procedure listed below:
	
	1. Generate a first-order reduced model that captures the phase dynamics of the oscillation. The first-order dynamics is described by the phase response curve (PRC). Perform PRC-based delay and advance entrainment on the full model, denote the corresponding light inputs as $u_{\rm delay}$ and $u_{\rm advance}$;
	
	2. Use the Principle Orthogonal Decomposition to obtain a second-order oscillator as a reduced model. Perform a direct shooting method \cite{Julius2018} to search the optimal light input for the minimum-time entrainment problem, denote it as $u_{\rm DSA}$;
	
	3. Perform a gradient descent algorithm to obtain the locally optimal light strategy $u^*$ on the full model. The gradient descent process is initialized by the results in steps 1, 2 and the 12-12-hour daily light-dark cycle $u_{\rm ref}$ in (\ref{eq:lis_ref}).

	\begin{figure}[thpb]
	\centering
	\includegraphics[width=3.2 in]{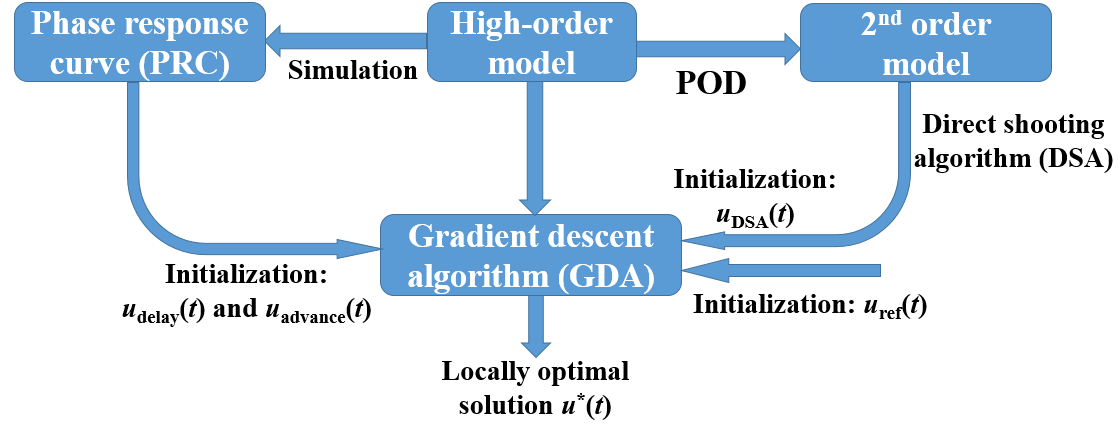}
	\caption{\small The solution procedure of the optimal light for the minimum-time entrainment of high-order models.}
	\label{fig:Solution_Procedure}
    \end{figure}		

	The circadian models we study in this paper are: A. the 7th-order ODE mammalian model in \cite{Geier2005}, B. the 3rd-order ODE Neurospora model in\cite{Leloup99}, C. the 10th-order ODE Drosophila model in\cite{Goldbeter98}. These models are introduced in Section II in details. In Section III, we introduce the theory and implementation of the solution procedure mentioned above. Section IV demonstrates the numerical implementation of entrainment strategies and simulation results. Section V draws some conclusions from the results.
	
	\section{Mathematic Model and Problem Formulation}
	
	\subsection{Mammalian Model}\label{Mammalian}
	
	Based on experimental observations on mice, computational mammalian circadian models were proposed in \cite{Sabine2004}\cite{Geier2005} as regulatory loops of $Per$, $Cry$, {\it Bmal1} genes transcription. The model in \cite{Geier2005} treats the $Per$ and $Cry$ genes as merely one variable. The 7 variables in this model are: the $Per/Cry$ mRNA ($y_1$), cytoplasmic PER/CRY complex ($y_2$), nucleus PER/CRY complex ($y_3$), cytoplasmic $Bmal1$ mRNA ($y_4$), cytoplasmic $Bmal1$ protein ($y_5$), nuclear $Bmal1$ ($y_6$) and modified nuclear $Bmal1$ ($y_7$). The dynamics of the states $y=[y_1,y_2,y_3,y_4,y_5,y_6,y_7]^T\in\mathbb{R}^{7\times1}$ is denoted as 
	\begin{equation}
	\dot y=F_{\rm mammal}(y,lis)=f_0(y)+f_1(y)lis,
	\label{eq:mammalian_model}
	\end{equation}
	where $lis$ is the light-related input of this model and valued between 0 and a constant value. In terms of \cite{Geier2005}, we set the maximum value of $lis$ is 0.02 in the presence of light. 
	
	\subsection{Neurospora Model}\label{Neurospora}
	
	The circadian rhythm in Neurospora regulates the periodic expression of circadian genes\cite{Tseng2012}. The circadian model of {\it Neurospora crassa} in\cite{Leloup99} is proposed based on the negative feedback regulation of the $frq$ gene transcription and FRQ protein synthesis. This model also takes the light effects into account, with the dynamics of this model denoted as
	\begin{equation}
	\frac{d[M, F_{\rm N}, F_{\rm C}]^T}{dt}=F_{\rm Neurospora}(M,F_{\rm N},F_{\rm C},v_{\rm s}),
	\label{eq:Neurospora_model}
	\end{equation}
	where $M$ is the $frq$ mRNA, $F_{\rm N}$ and $F_{\rm C}$ are FRQ protein in the nucleus and cytoplasm, respectively. $v_{\rm s}$ denotes the $frq$ transcriptional rate, which is a light-dependent variable, we set $v_{\rm s}=1.6$ in dark and $v_{\rm s}=2$ in full light, i.e., $v_{\rm s}\in[1.6,2]$. 
	
	\subsection{Drosophila Model}\label{Drosophila}
	
	Similar to the Neurospora model, the Drosophila circadian model is formulated based on the negative feedback regulation of $per$ and $tim$ genes transcription and PER and TIM proteins synthesis\cite{Goldbeter98}\cite{Leloup99}. The dynamics of the 10 states of the Drosophila model in \cite{Goldbeter98} are denoted as
	\begin{gather}
	\frac{d}{dt}[M_{\rm P},P_0,P_1,P_2,M_{\rm T},T_0,T_1,T_2,C,C_{\rm N}]^T\nonumber\\=F_{\rm Drosophila}(M_{\rm P},P_0,P_1,P_2,M_{\rm T},T_0,T_1,T_2,C,C_{\rm N},v_{\rm dT}),
	\label{eq:Drosophila_model}
	\end{gather}
	where $M_{\rm P}$ and $M_{\rm T}$ represent the $per$ and $tim$ mRNA, $P_0, P_1, P_2$ represent three different states of the PER protein,  $T_0, T_1, T_2$ represent three different states of the TIM protein, $C$ and $C_N$ represent the PER-TIM complex in the cytoplasm and nucleus, respectively. The total quantities of the PER and TIM proteins are given by:
	$$P_t=P_0+P_1+P_2+C+C_N,\,\,T_t=T_0+T_1+T_2+C+C_N.$$
	The maximum rate of TIM protein degradation $v_{\rm dT}$ is a light-dependent variable. Based on \cite{Goldbeter98}, $v_{\rm dT}$ is valued at 2 and 4 in dark and light, respectively.
	
	\subsection{Problem Formulation}\label{Problem}
	
	Assume the light-dependent variable in each circadian model is expressed as $u\in\mathbb{R}^1$, which is used to represent the light input. In the minimum-time entrainment problem, we want to find the optimal light $u^*(t)\in[u_{\rm min},u_{\rm max}]$ to drive the circadian state $x(t)\in\mathbb{R}^{n}$ to a reference trajectory $x_{\rm ref}(t)$ in minimum time given the initial condition $x(0)$ and the system dynamics
	\begin{equation} 
	\dot x=F(x(t),u(t)),
	\label{system_dynamics}
	\end{equation} 
	where $n$ is the number of model states. The system dynamics of each model are given as (\ref{eq:mammalian_model}), (\ref{eq:Neurospora_model}) and (\ref{eq:Drosophila_model}), $u_{\rm min}$ and $u_{\rm max}$ are the upper and lower bounds of the light input. At the final time $t_f$, the circadian state should satisfy
	\begin{gather}
	\varphi_f(t_f,x(t_f))=||x(t_f)-x_{\text{ref}}(t_f)||^2-tol=0,\label{stopping_criterion}
	\end{gather}
	where $tol$ is valued at a small positive scalar and represents the final error between the entraining and reference state.
	
	\section{Solution Process and Entrainment Strategy}\label{Strategy}
	
	\subsection{\textbf{Open-loop Entrainment}}\label{Openloop}
	
	The most convenient and widely-used entrainment strategy is applying the daily light-dark cycle on the subject directly. In each model, we assume the reference light $u_{\rm ref}$ is the 12/12 hours light/dark cycle, with a light input given as
	\begin{equation}
	u_{\rm ref}(t) = \left \{ 
	\begin{array}{cl}
	u_{\rm max}, & {\rm mod}(t,24)\in[0,12),\\
	u_{\rm min}, & {\rm mod}(t,24)\in[12, 24).
	\end{array}
	\right.
	\label{eq:lis_ref}
	\end{equation}
	The dynamics of the circadian state in this entrainment process is given as
	\begin{gather}
	\dot x=F(x(t),u_{\rm ref}(t)).\label{openloop_entrainment}
	\end{gather}
	The reference trajectory $x_{\rm ref}(t)$ is set as the periodic solution of the differential equation in (\ref{openloop_entrainment}). The light $u(t)=u_{\rm ref}(t)$ is a function of time and does not depend on the circadian state, we call the entrainment by $u_{\rm ref}(t)$ as {\it open-loop entrainment}. 
	
	\subsection{\textbf{PRC and Greedy Delay/Advance Entrainment}}\label{SVD}
	
	PRC is widely used in the study of circadian rhythm \cite{Sacre2014} and design of light input for circadian entrainment\cite{Qiao2017}. In this paper, the circadian phase of each model is defined by
	\begin{equation}
	\theta = \left \{ 
	\begin{array}{cl}
	{\rm tan}^{-1}(\bar{y}_2/\bar{y}_1), & {\rm in\ mammalian\ model},\\
	{\rm tan}^{-1}(\bar{F}_{\rm C}/\bar{M}), & {\rm in\ Neurospora\ model},\\
	{\rm tan}^{-1}(\bar{C}_{\rm N}/\bar{M}_{\rm T}), & {\rm in\ Drosophila\ model},
	\end{array}
	\right.
	\label{eq:circadian_phase}
	\end{equation}
	where $\bar{X}$ represents the normalized state $X$, given as $\bar{X}=[2X-(X_{\rm max}+X_{\rm min})]/(X_{\rm max}-X_{\rm min})$, $X_{\rm min}$ and $X_{\rm max}$ are the minimum and maximum values of $X$. The dynamics of the circadian phase is given as $\dot\theta=\omega_0(\theta)+f(\theta)u$, where $\omega_0(\theta)$ is the free running frequency (the frequency in dark). $f(\theta)$ is the phase response curve, which is determined by adding a 30-min light pulse on the free running state (circadian state in dark) and calculating the phase differences with and without light. The results in \cite{Qiao2017} imply that the optimal control for circadian phase entrainment is either the greedy delay light $u(t)=(u_{\rm max}-u_{\rm min})[1-{\rm sign}(f(\theta))]/2+u_{\rm min}$ or greedy advance light $u(t)=(u_{\rm max}-u_{\rm min})[1+{\rm sign}(f(\theta))]/2+u_{\rm min}$. Note that the amplitude of the circadian state can be quenched or enlarged by the light input. To improve the convergence property, we add a part of reference light as
    \begin{equation}
    u_{\rm delay}(\theta,\theta_{\rm ref}) = \left \{ 
    \begin{array}{cl}
    u_{\rm max}, & f(\theta)<0\ {\rm and}\ \Delta\theta\geq\Upsilon,\\
    u_{\rm min}, & f(\theta)\geq 0\ {\rm and}\ \Delta\theta\geq\Upsilon,\\
    u_{\rm ref}(\theta_{\rm ref}), & \Delta\theta<\Upsilon,
    \end{array}
    \right.
    \label{eq:u_delay_revised}
    \end{equation}
    \begin{equation}
    u_{\rm advance}(\theta,\theta_{\rm ref}) = \left \{ 
    \begin{array}{cl}
    u_{\rm max}, & f(\theta)>0\ {\rm and}\ \Delta\theta\geq\Upsilon,\\
    u_{\rm min}, & f(\theta)\leq 0\ {\rm and}\ \Delta\theta\geq\Upsilon,\\
    u_{\rm ref}(\theta_{\rm ref}), & \Delta\theta<\Upsilon,
    \end{array}
    \right.
    \label{eq:u_advance_revised}
    \end{equation}
    where the phase difference $\Delta\theta$ is defined as $\Delta\theta=\min\{|\theta-\theta_{\rm ref}\pm2\pi|,|\theta-\theta_{\rm ref}|\}$, and $\Upsilon=1$ rad. Since the reference trajectory keeps synchronized with $u_{\rm ref}(t)$, we represent $u_{\rm ref}(t)$ as $u_{\rm ref}(\theta_{\rm ref})$ in (\ref{eq:u_delay_revised}) and (\ref{eq:u_advance_revised}). Assuming the circadian phase of the full model is available during entrainment, we can implement $u_{\rm delay}\ {\rm and}\ u_{\rm advance}$ as feedback control laws.
	
	\subsection{\textbf{Model-Order Reduction and Direct Shooting Algorithm}}\label{2nd}
	The direct shooting algorithm (DSA) has been introduced in \cite{Julius2018} and \cite{Zhang2012} to solve the minimum-time entrainment problem and search for a \emph{globally optimal} solution in the second-order Kronauer model. However, this method is \emph{impractical for models of third or higher orders}. Therefore,  before running the shooting algorithm, we reduce the full model into a second-order one. Assume the dynamics of the high-order model is given as (\ref{system_dynamics})	with the initial condition $x(0)=x_0$. The approximation of state using a right reduced-order base $U_r\in\mathbb{R}^{n\times2}$ is given as
	$$x(t,u(t);x_0)\approx U_rz(t,u(t);z_0).$$
	If $U_r^TU_r=I$, the reduced nonlinear model is given as
	\begin{equation}
	\dot z(t)\approx U_r^TF(U_rz(t,u(t);z_0),u(t)),\ \text{where}\ z_0\approx U_r^Tx_0.
	\label{eq:reduced_model}
	\end{equation}
	The projection matrix $U_r$ of a nonlinear model is usually generated by Principal Orthogonal Decomposition\cite{Lenaerts2001}, which finds a subspace in $\mathbb{R}^2$ that captures information of the original model by minimizing the $\mathcal{L}_2$ norm of error $||x(t)-U_rz(t)||_2$. We simulate the circadian model forward in darkness and generate a free-running trajectory $x_{\rm free}(t)$ in several periods, set a sequence of sampling time $\{t_1,t_2,...,t_N\}$ and construct a {\it snapshot matrix} of the trajectory, denoted as 
	$$X_{\rm free}=\{x_{\rm free}(t_1)\ x_{\rm free}(t_2)\ ...\ x_{\rm free}(t_N)\}\in\mathbb{R}^{n\times N}.$$
	By singular value decomposition (SVD), this matrix is decomposed in the following form:
	$$X_{\rm free}=U\Sigma V^*=[u_1\ u_2...u_n]\Sigma V^*\approx U_r\Sigma_rV^*_r=[u_1\ u_2]\Sigma_rV_r^*,$$
	where $U_r\in\mathbb{R}^{n\times 2},\ \Sigma_r\in\mathbb{R}^{2\times2},\ V_r\in\mathbb{R}^{N\times2}$ are truncated SVD matrices which contain the first two principle columns and singular values. 
	
	We use the optimal control theory\cite{Zhang2012} to solve the minimum-time entrainment problem in \ref{Problem} with a reduced model in (\ref{eq:reduced_model}). The cost function is usually expressed in an integral form as $J=t_f=\int_{0}^{t_f} 1d\tau$.
	The Hamiltonian of the reduced system is given as $\mathscr{H}=1+p^T\dot z$, where $p(t)\in\mathbb{R}^{2}$ is the Lagrange multiplier and is usually called the {\it co-state}. The dynamics of the co-state is given as
	\begin{equation}
	\dot p=-\frac{\partial\mathscr{H}}{\partial z}=-\frac{\partial\dot z}{\partial z}p.
	\label{eq:costate_equation}
	\end{equation}
	Based on the Pontryagin Minimum Principle, the optimal control $u^*$ satisfies
	\begin{equation}
	u^*(t)=\arg\min_{u(t)}\mathscr{H}=\frac{u_{\rm max}-u_{\rm min}}{2}[1-{\rm sign}(p^T\dot z)]+u_{\rm min}.
	\label{eq:optimal_u}
	\end{equation}
	The terminal co-state is decided by transversality condition
	\begin{equation}
	p(t_f)=-\frac{\frac{\partial\varphi_f(t_f,z(t_f))}{\partial z(t_f)}}{\frac{\partial\varphi_f(t_f,z(t_f))}{\partial t_f}+\left[\frac{\partial\varphi_f(t_f,z(t_f))}{\partial z(t_f)}\right]^T\dot z(t_f)}.
	\label{eq:transversality_condition}
	\end{equation}
	The minimum-time entrainment problem is treated as a two-point boundary values problem for solving the optimal initial co-state value $p(0)\in\mathbb{R}^{2}$ given as $p^*(0)=\arg\min_{p(0)} t_f.$
	
	The solution is subject to the initial condition $z(0)=z_0$, the terminal condition $\varphi_f(t_f,z(t_f))=0$ and optimality constraints (\ref{eq:costate_equation}), (\ref{eq:optimal_u}), (\ref{eq:transversality_condition}).
	Note that the magnitude of $p(0)=[p_1(0),p_2(0)]^T$ has no effect on the entrainment process and the final result. Thus, we limit our search on a unit circle as $p(0)=[\cos(\phi),\sin(\phi)]^T$. The true magnitude of $p(t_f)$ can later be determined by (\ref{eq:transversality_condition}). The direct shooting algorithm only need to search the optimal value of $\phi\in[0,2\pi]$ with a process given in \cite{Zhang2012}. The light input $u(t)$ corresponding to $p^*(0)$ is the optimal light input for minimum-time entrainment of the reduced model, denoted as $u_{\rm DSA}(t)$. We apply $u_{\rm DSA}(t)$ on the full model in the form of (\ref{eq:u_2nd}), where $t_{\rm DSA}$ is the optimal entrainment time of the reduced model.
	
	\begin{equation}
	u_{\rm 2nd}(t) = \left \{ 
	\begin{array}{cl}
	u_{\rm DSA}(t), & t\leq t_{\rm DSA},\\
	u_{\rm ref}(t), & {\rm otherwise}.
	\end{array}
	\right.
	\label{eq:u_2nd}
	\end{equation}
	
	\subsection{\textbf{Gradient Descent Algorithm}}\label{Optimal}
	
	A gradient descent algorithm is used for solving the minimum-time entrainment problem of the full model. The system dynamics and the terminal constraint are plugged into the cost function by introducing multipliers $\lambda(t)\in\mathbb{R}^{n\times1}$ and $\epsilon\in\mathbb{R}^1$\cite{Bryson1975}, the augmented cost function is written as
	\begin{gather}
	J_a(u)=\int_{0}^{t_f}1+\lambda^T(\tau)[F(x(\tau),u(\tau))-\dot x(\tau)]d\tau\nonumber\\+\epsilon\varphi_f(t_f,x(t_f)).\label{augmented_cost}
	\end{gather}
	For the feasible state $x(t)$ and corresponding time cost $t_f$ given $x_0$ and $u(t)$, we introduce a small perturbation into $u$, the perturbed light input is given as $u(t)+\alpha\xi(t)$, where $\alpha$ is a scalar. The cost with variation in the light input is given as $J_a(u+\alpha\xi)$. Note that $\lambda(t)$ and $\epsilon$ have no effects on the cost value. For simplification, $\lambda(t)$ and $\epsilon$ are defined as:
	\begin{gather}
	\dot\lambda(t)=-\lambda^T\frac{\partial F(x(t),u(t))}{\partial x},\lambda(t_f)=\epsilon\frac{\partial\varphi_{f}(t_f,x(t_f))}{\partial x},\label{co-state}\\
	\epsilon=-\frac{1}{\frac{\partial\varphi_{f}(t_f,x(t_f))}{\partial t}+\frac{\partial\varphi_{f}(t_f,x(t_f))}{\partial x}F(x(t_f),u(t_f))}.\label{co-state3}
	\end{gather}
	The first order variation of cost with respect to light is
	\begin{gather}
	\delta J_a|_u=\lim_{\alpha\rightarrow 0}
	\frac{\partial [J_a(u+\alpha\xi)-J_a(u)]}{\partial\alpha}\nonumber\\=\int_{0}^{t_f}\lambda^T(\tau)\frac{\partial F(x(\tau),u(\tau))}{\partial u}\xi(\tau)d\tau.\label{cost_variation}
	\end{gather}
	As $J_a=J$, the gradient descent direction of the light is 
	\begin{equation}
	\nabla_{u(t)}J=\lambda^T(t)\frac{\partial F(x(t),u(t))}{\partial u}.\label{gradient_u}%
	\end{equation}
	Steps of the gradient descent process for solving the optimal light input of the minimum-time circadian entrainment problem are listed below:
	\begin{algorithm}[H]
		\begin{algorithmic}[1]
			\caption{Gradient descent algorithm (GDA):}
			\State {\small choose an initial guess of the light input $u^{0}(t)$, $k\gets0$};
			\While{\small $k=0\ {\rm or}\ u_{k}\neq u_{k-1}$}
			\State {\small integrate the state equation $\dot x=F(x(t),u^k(t))$ forward until $\varphi_f=0$ and determine the final time $t^k_f$ and final state $x^k(t^k_f)$};
			\State {\small Determine the final value of $\lambda^k(t^k_f)$ and simulate backward to get $\lambda^k(t),\ \forall t\in[0,t^k_f]$} based on (\ref{co-state});	
			\State {\small Determine the gradient descent direction of $u^k(t)$ based on (\ref{gradient_u}) , update $u^k$ by
				\begin{equation}
				\small u^{k+1}(t)=\min\left\{\max\left[u^{k}(t)-\eta_u\nabla_{u^k(t)}J,u_{\rm min}\right],u_{\rm max}\right\},\label{Update_u}%
				\end{equation}
				where the updating step $\eta_u=\arg\min_{\eta>0}J(u^{k}(t)-\eta\nabla_{u^k(t)}J)$ is solved by a line search;}
			\State {\small $k\gets k+1$;} 
			\EndWhile
			\State{\small $u^*(t)=u^{k}(t)$.}
			\label{GDA}
		\end{algorithmic}
	\end{algorithm}	
	{\it Lemma}: The gradient descent process reaches a fixed point $u^*(t)=u^{k+1}(t)=u^{k}(t)$ for any positive step $\eta_u$ if and only if the following conditions are satisfied for $\forall t\in[0,t_f^k]$:
	
	\begin{equation}
	\left\{\begin{array}
	[c]{c}%
	u^{k}(t)=u_{\rm min}\ {\rm when}\ \nabla_{u^k(t)}J>0,\\
	u^{k}(t)=u_{\rm max}\ {\rm when}\ \nabla_{u^k(t)}J<0,\\
	u^{k}(t)\in[u_{\rm min},u_{\rm max}]\ {\rm when}\ \nabla_{u^k(t)}J=0.
	\end{array}
	\right.\label{eq:u}
	\end{equation}
	The result in \textbf{Algorithm \ref{GDA}} relies heavily on the initial guess. To get a locally optimal solution with small entrainment time, we choose four initial guesses for GDA: a) $u^0(t)=u_{\rm ref}(t)$, b) $u^0(t)=u_{\rm advance}(t)$, c) $u^0(t)=u_{\rm delay}(t)$, d) $u^0(t)=u_{\rm 2nd}(t)$.
	
	\section{Numerical Implementation}\label{numerical}
	
	Consider the case that the subject is entrained under the reference light at the beginning. A sudden time shift $\Delta_{\rm shift}$ in the external light-dark cycle occurs at $t=0$ and the subject loses synchronization with the local environment, that is, $x(0)=x_{\rm ref}(\Delta_{\rm shift})\neq x_{\rm ref}(0)$, $\Delta_{\rm shift}\in\{1,2,...,23\}$. We want to re-entrain the subject as quickly as possible.
	
	\subsection{Mammalian Model}
	
\begin{figure}[H]
		\centering
		\begin{subfigure}[b]{0.233\textwidth}
			\includegraphics[width=\textwidth]{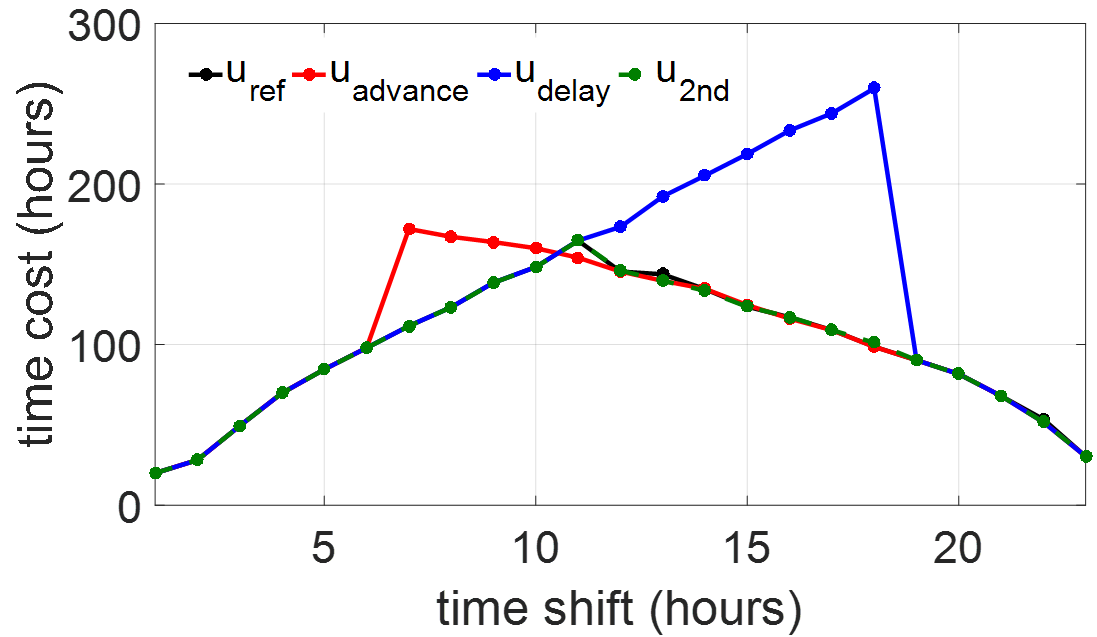}
			\caption{\small GDA results}
			\label{fig:Mammalian_Entrainment_initial}
		\end{subfigure}
		\begin{subfigure}[b]{0.245\textwidth}
			\includegraphics[width=\textwidth]{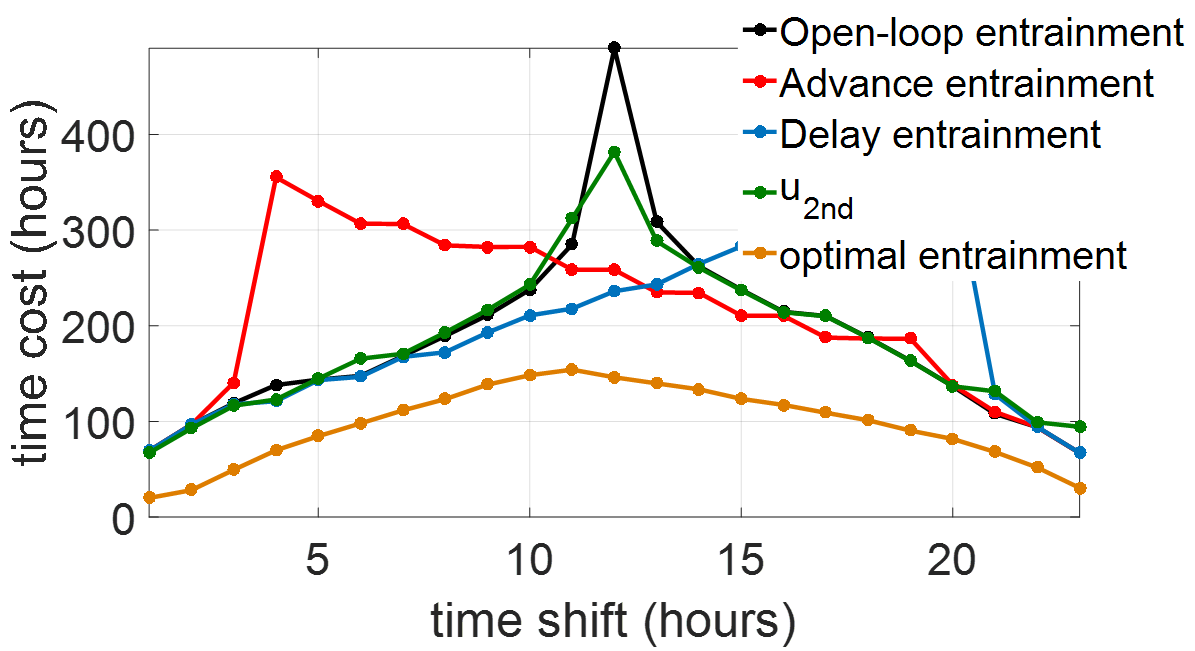}
			\caption{\small Entrainment time}
			\label{fig:Mammalian_Entrainment_Time}
		\end{subfigure}
		\caption{\small The gradient descent results and entrainment time cost of open-loop entrainment, delay/advance entrainment, entrainment of $u_{\rm 2nd}$ and optimal entrainment of the mammalian model, the optimal entrainment time is the minimum time of the gradient descent results among four initial guesses.}
		\label{fig:Mammalian_Model}  
	\end{figure}

	For the mammalian model, the circadian state $x(t)=y(t)$ and the light input $u=lis$. Fig. \ref{fig:Mammalian_Model} shows the gradient descent results with four initial guesses (in Fig. \ref{fig:Mammalian_Model}a) and various entrainment time costs (in Fig. \ref{fig:Mammalian_Model}b). The PRC-based delay-advance threshold occurs at about 13 hours shift, the delay/advance entrainment works better than $u_{\rm ref}$ and $u_{\rm 2nd}$ in the cases with a time shift of 8-16 hours. The GDA results show that different initial lights may converge to different locally optimal solutions. In the 11 hours shift case, the GDA initialized by $u_{\rm advance}(t)$ reduces the time cost from 260 to 154 hours, while that initialized by $u_{\rm delay}(t)$ decreases the cost from 210 to 164 hours. 
	
\begin{figure}[htbp]
	\centering
	\begin{subfigure}[b]{0.20\textwidth}
		\includegraphics[width=\textwidth]{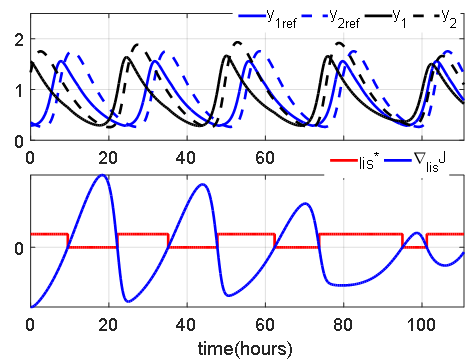}
		\caption{\small 8 hours shift}
		\label{fig:8_shift_Mammalian_y12}
	\end{subfigure}
	\begin{subfigure}[b]{0.20\textwidth}
		\includegraphics[width=\textwidth]{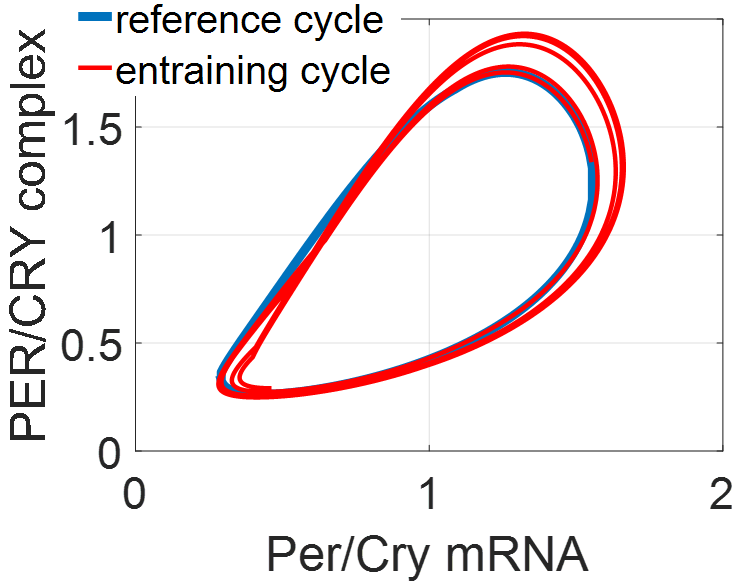}
		\caption{\small $Per$-$Cry$ mRNA-protein}
		\label{fig:8_shift_Mammalian_limit_cycle_mRNA}
	\end{subfigure}
	\begin{subfigure}[b]{0.20\textwidth}
		\includegraphics[width=\textwidth]{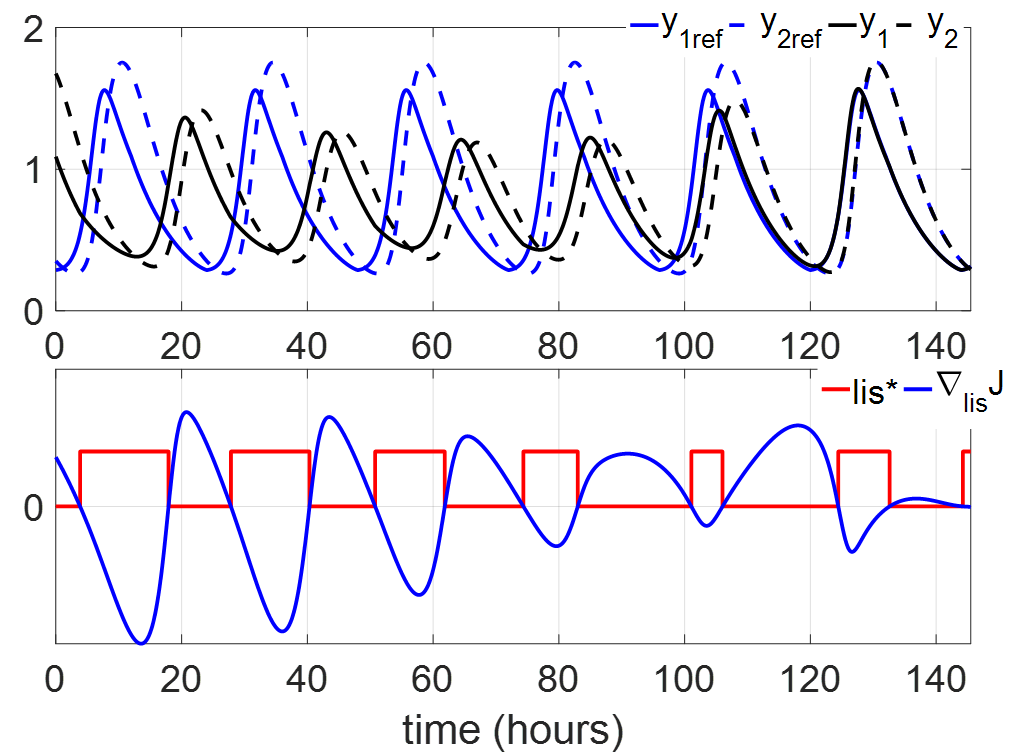}
		\caption{\small 12 hours shift}
		\label{fig:12_shift_Mammalian_y12} 
	\end{subfigure}
	\begin{subfigure}[b]{0.20\textwidth}
		\includegraphics[width=\textwidth]{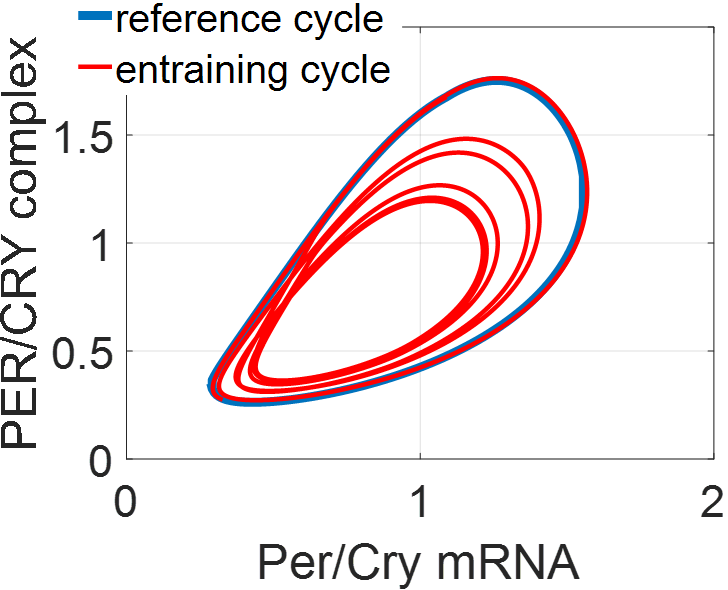}
		\caption{\small $Per$-$Cry$ mRNA-protein}
		\label{fig:12_shift_Mammalian_limit_cycle_mRNA}
	\end{subfigure}
	\begin{subfigure}[b]{0.20\textwidth}
		\includegraphics[width=\textwidth]{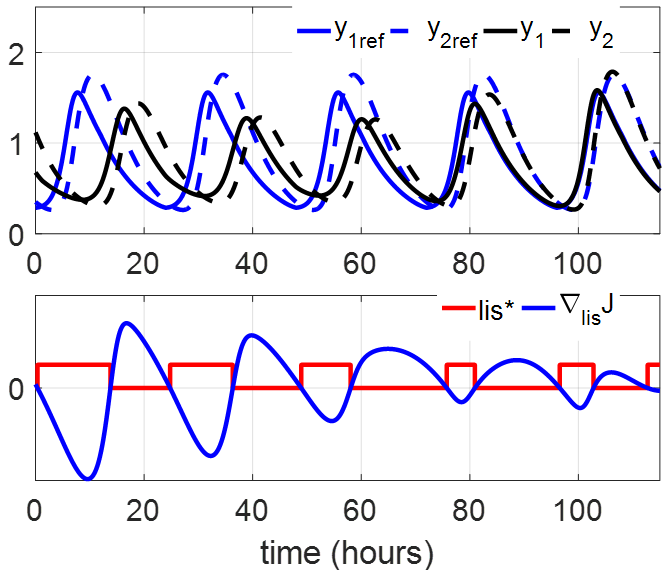}
		\caption{\small 16 hours shift}
		\label{fig:16_shift_Mammalian_y12}
	\end{subfigure}
	\begin{subfigure}[b]{0.20\textwidth}
		\includegraphics[width=\textwidth]{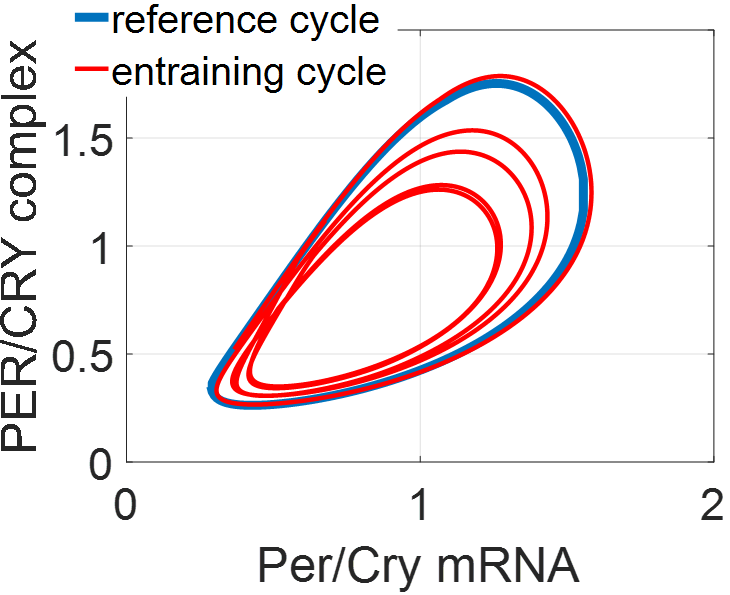}
		\caption{\small $Per$-$Cry$ mRNA-protein}
		\label{fig:16_shift_Mammalian_limit_cycle_mRNA}
	\end{subfigure}  
	\caption{\small Optimal entrainment cases of the mammalian model.}
	\label{fig:Entrainment_cases_Mammalian2}  
\end{figure}

Fig. \ref{fig:Entrainment_cases_Mammalian2} shows 8, 12 and 16-hour shifts cases by the optimal light and time evolution of the $Per/Cry$ mRNA and protein during the optimal entrainment process. The lower sub-figures in Fig. \ref{fig:8_shift_Mammalian_y12}, \ref{fig:12_shift_Mammalian_y12}, \ref{fig:16_shift_Mammalian_y12} compare the gradient $\nabla_{lis}J$ with the optimal light $lis^*(t)$, implying that the light $lis^*(t)$ in these figures follows the optimality condition mentioned in (\ref{eq:u}). Note that the optimal controls in these cases are all bang-bang control, i.e., $lis^*(t)$ is either maximum or minimum. Fig. \ref{fig:8_shift_Mammalian_limit_cycle_mRNA}, \ref{fig:12_shift_Mammalian_limit_cycle_mRNA} and \ref{fig:16_shift_Mammalian_limit_cycle_mRNA} show the $Per/Cry$ mRNA--PER/CRY protein cycle during the optimal entrainment process. Define the time shift corresponding to the maximum time cost in the optimal entrainment as $\Delta_{\rm threshold}$. In this mammalian model, $\Delta_{\rm threshold}\approx11$ hours, we can seen that when $\Delta_{\rm shift}>\Delta_{\rm threshold}$, the amplitude of $Per$-$Cry$ mRNA--protein oscillator is quenched during entrainment, i.e., for 12 and 16 hours shifts; while in the case with 8 hours shift, the amplitude of mRNA-protein cycle is enlarged, the optimal light tend to enhance the $Per/Cry$ mRNA transcription and PER/CRY protein synthesis in this case.

	\subsection{Neurospora Model}

\begin{figure}[htbp]
	\centering
	\begin{subfigure}[b]{0.235\textwidth}
		\includegraphics[width=\textwidth]{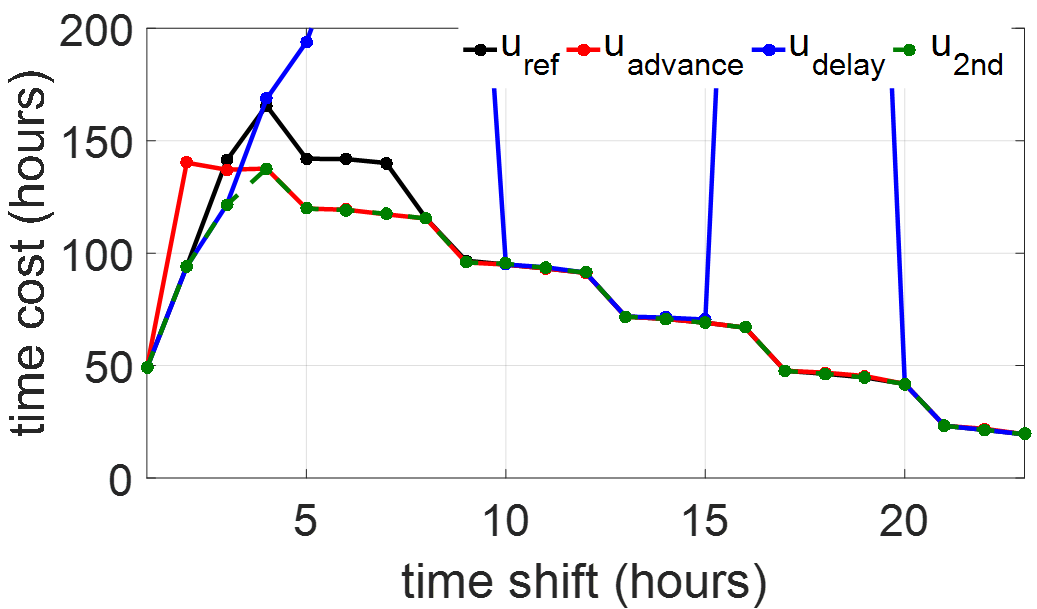}
		\caption{\small GDA results}
		\label{fig:Neurospora_Entrainment_initial}
	\end{subfigure}
	\begin{subfigure}[b]{0.235\textwidth}
		\includegraphics[width=\textwidth]{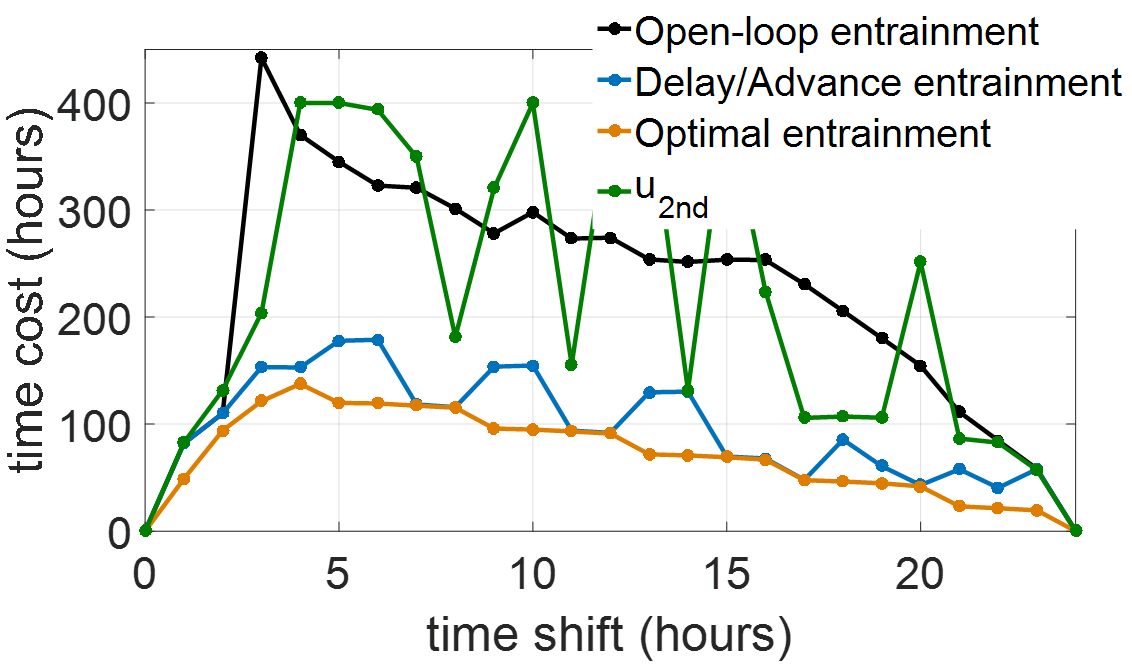}
		\caption{\small Entrainment time}
		\label{fig:Neurospora_Entrainment_Time}
	\end{subfigure}
	\caption{\small The GDA results and entrainment time of the Neurospora model, in which the delay/advance entrainment time is the minimum entrainment time among the delay and advance strategies.}
	\label{fig:Neurospora_Model}  
    \end{figure}

	In the Neurospora model, $x=[M,F_{\rm C},F_{\rm N}]^T$ and $u(t)=v_{\rm s}(t)$. Fig. \ref{fig:Neurospora_Entrainment_initial} shows the GDA results in the Neurospora model, we can observe that the DSA results from the reduced models are the best initial guess of the GDA as the GDA initialized by $u_{\rm 2nd}$ always reaches the best solution among all four initial guesses in every case. Fig. \ref{fig:Neurospora_Entrainment_Time} demonstrates the entrainment time of four strategies in the Neurospora model. Compared with the open-loop entrainment and entrainment by $u_{\rm 2nd}$, the delay/advance strategy and optimal light dramatically decrease the time cost, especially in cases with $\Delta_{\rm shift}\in[3,20]$ hours. The maximum time cost in the optimal entrainment occurs at $\Delta_{\rm threshold}=$4 hours shift. This implies the Neurospora model is an asymmetrical one in which advancing 18 hours is faster than delaying 6 hours.

\begin{figure}[H]
		\centering
		\begin{subfigure}[b]{0.20\textwidth}
			\includegraphics[width=\textwidth]{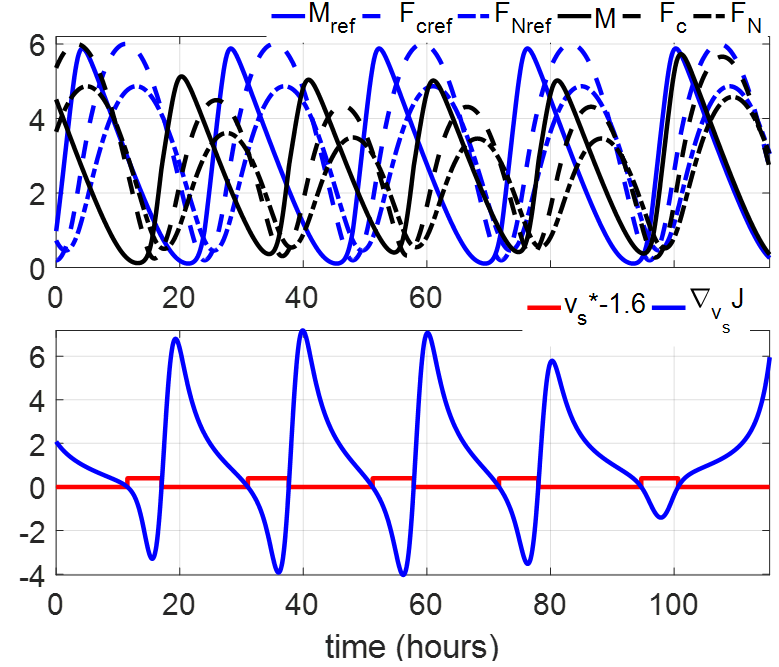}
			\caption{\small 8 hours shift}
			\label{fig:8_shift_Neurospora}
		\end{subfigure}
		\begin{subfigure}[b]{0.20\textwidth}
			\includegraphics[width=\textwidth]{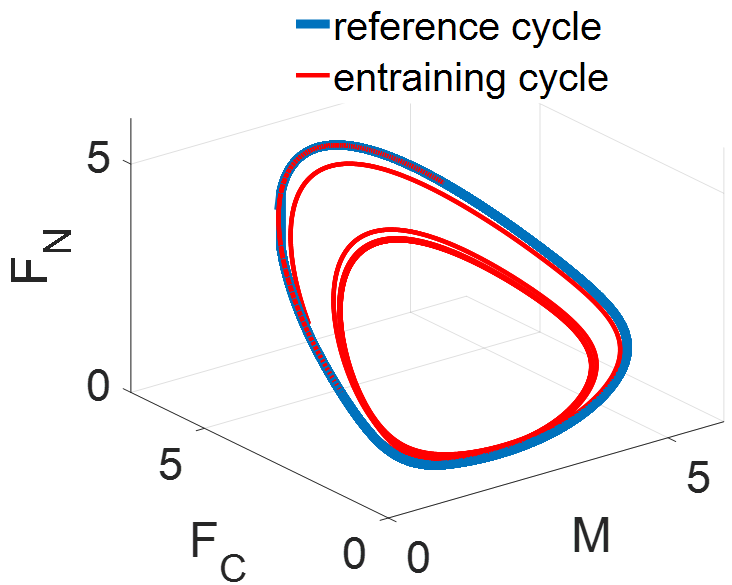}
			\caption{\small $M$-$F_C$-$F_N$ cycle}
			\label{fig:8_shift_Neurospora_limit_cycle}
		\end{subfigure}
		\begin{subfigure}[b]{0.20\textwidth}
			\includegraphics[width=\textwidth]{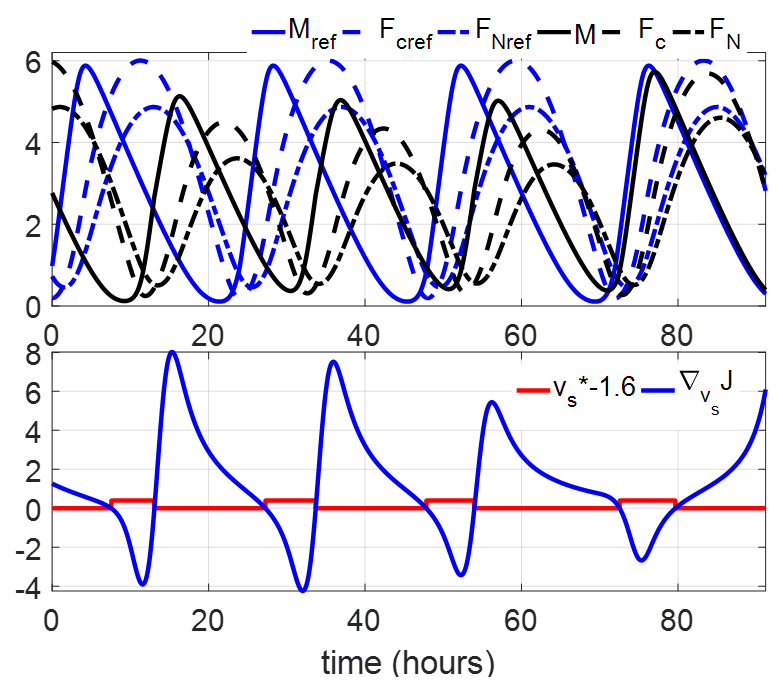}
			\caption{\small 12 hours shift}
			\label{fig:12_shift_Neurospora} 
		\end{subfigure}
		\begin{subfigure}[b]{0.20\textwidth}
			\includegraphics[width=\textwidth]{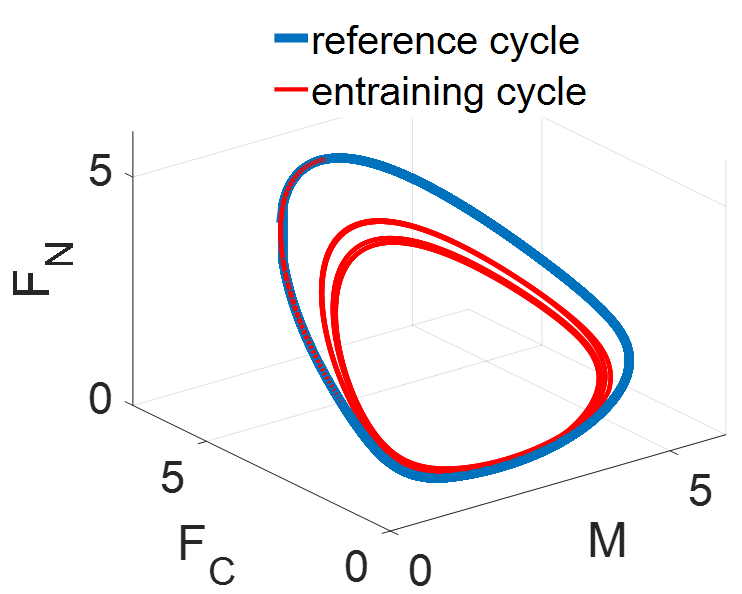}
			\caption{\small $M$-$F_C$-$F_N$ cycle}
			\label{fig:12_shift_Neurospora_limit_cycle}
		\end{subfigure}
		\begin{subfigure}[b]{0.20\textwidth}
			\includegraphics[width=\textwidth]{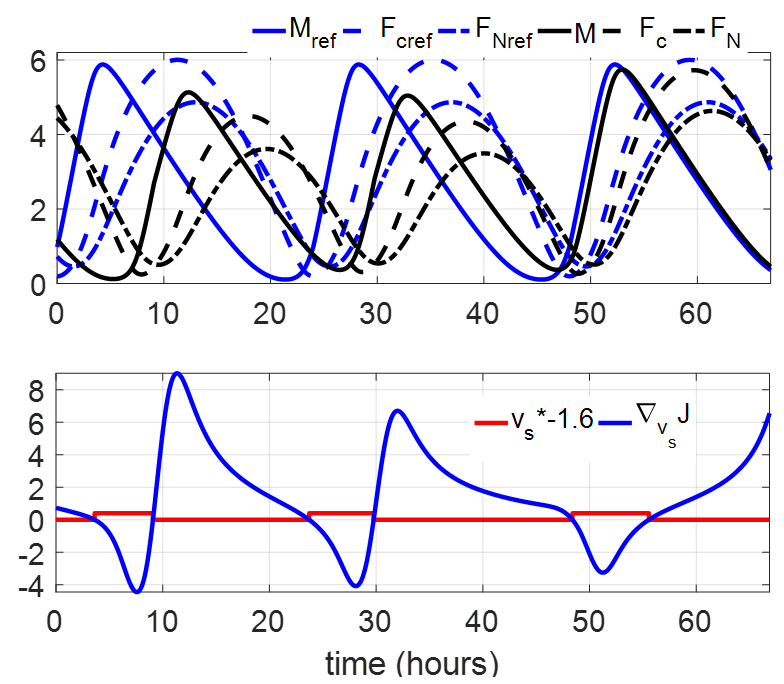}
			\caption{\small 16 hours shift}
			\label{fig:16_shift_Neurospora}
		\end{subfigure}
		\begin{subfigure}[b]{0.20\textwidth}
			\includegraphics[width=\textwidth]{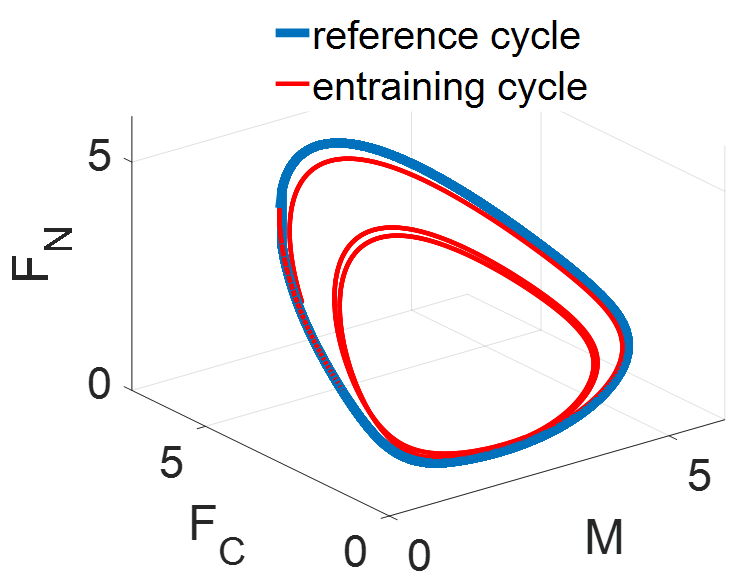}
			\caption{\small $M$-$F_C$-$F_N$ cycle}
			\label{fig:16_shift_Neurospora_limit_cycle}
		\end{subfigure}  
		\caption{\small Optimal entrainment cases of the Neurospora model.}
		\label{fig:Entrainment_cases_Neurospora}  
	\end{figure}

    Fig. \ref{fig:Entrainment_cases_Neurospora} plots 8, 12 and 16-hour shifts optimal entrainment cases of the Neurospora model. Similar to the results in the mammalian model, the optimal control in the Neurospora model is a bang-bang control. From the three cases ($\Delta_{\rm shift}>\Delta_{\rm threshold}$) shown in Fig. \ref{fig:Entrainment_cases_Neurospora}, we observe that during the optimal entrainment process, the amplitudes of the Neurospora oscillator are all quenched, the optimal light decreases the rate of $frq$ mRNA transcription and FRQ protein synthesis.

	\subsection{Drosophila Model}

\begin{figure}[htbp]
		\centering
		\begin{subfigure}[b]{0.235\textwidth}
			\includegraphics[width=\textwidth]{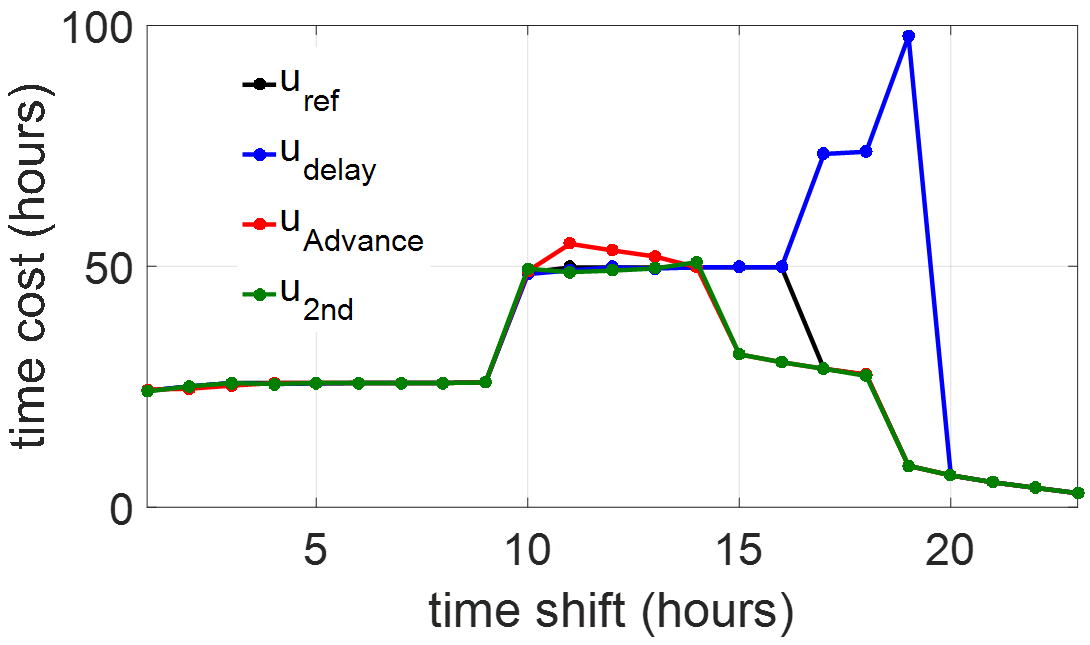}
			\caption{\small GDA results}
			\label{fig:Leloup_Entrainment_initial}
		\end{subfigure}
		\begin{subfigure}[b]{0.235\textwidth}
			\includegraphics[width=\textwidth]{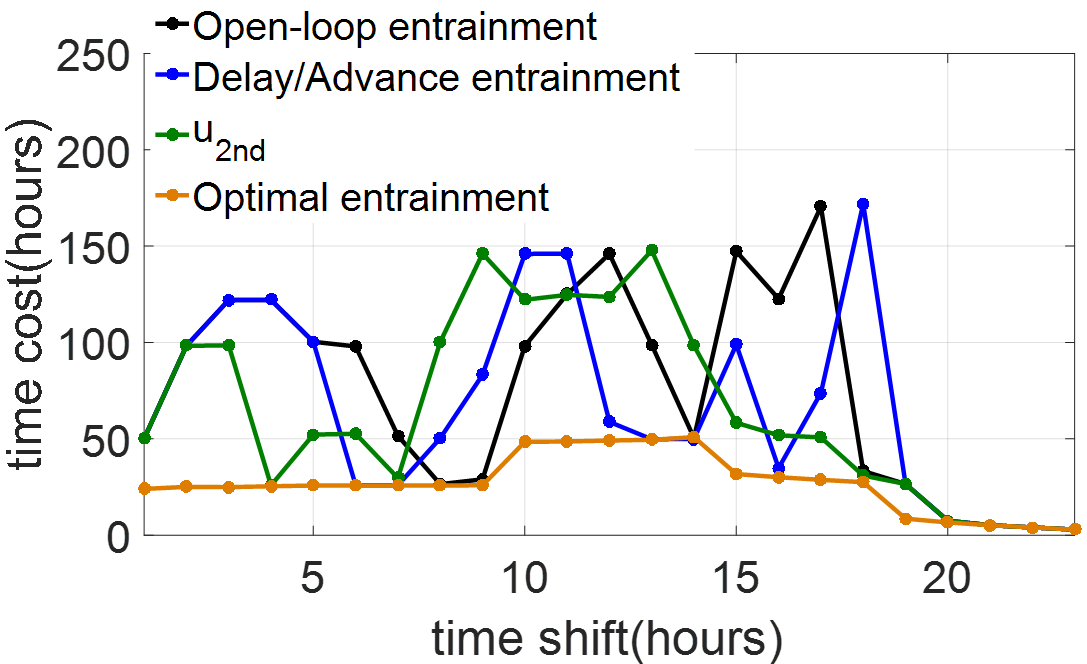}
			\caption{\small Entrainment time}
			\label{fig:Leloup_Entrainment_Time}
		\end{subfigure}
		\caption{\small GDA results and entrainment time of the Drosophila model.}
		\label{fig:Leloup_Model}  
	\end{figure}

\begin{figure}[htbp]
		\centering
		\begin{subfigure}[b]{0.20\textwidth}
			\includegraphics[width=\textwidth]{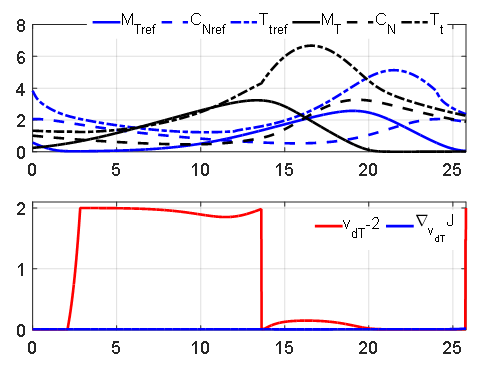}
			\caption{\small 8 hours shift}
			\label{fig:8_shift_Leloup2}
		\end{subfigure}
		\begin{subfigure}[b]{0.19\textwidth}
			\includegraphics[width=\textwidth]{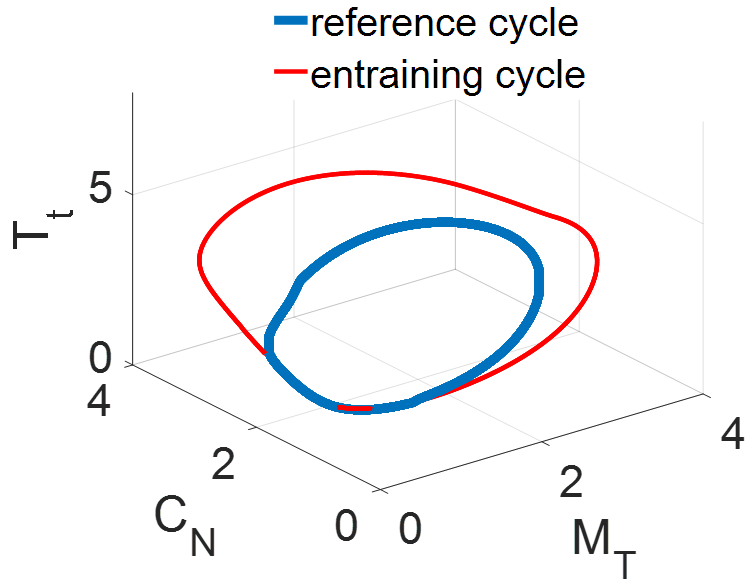}
			\caption{\small $M_{\rm T}$-$C_{\rm N}$-$T_t$ cycle}
			\label{fig:8_shift_Leloup_MT_CN_Tt}
		\end{subfigure}
		\begin{subfigure}[b]{0.20\textwidth}
			\includegraphics[width=\textwidth]{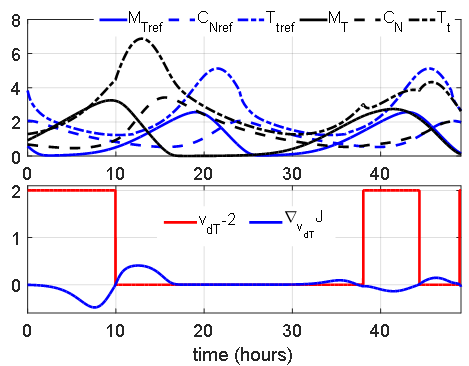}
			\caption{\small 12 hours shift}
			\label{fig:12_shift_Leloup2} 
		\end{subfigure}
		\begin{subfigure}[b]{0.19\textwidth}
			\includegraphics[width=\textwidth]{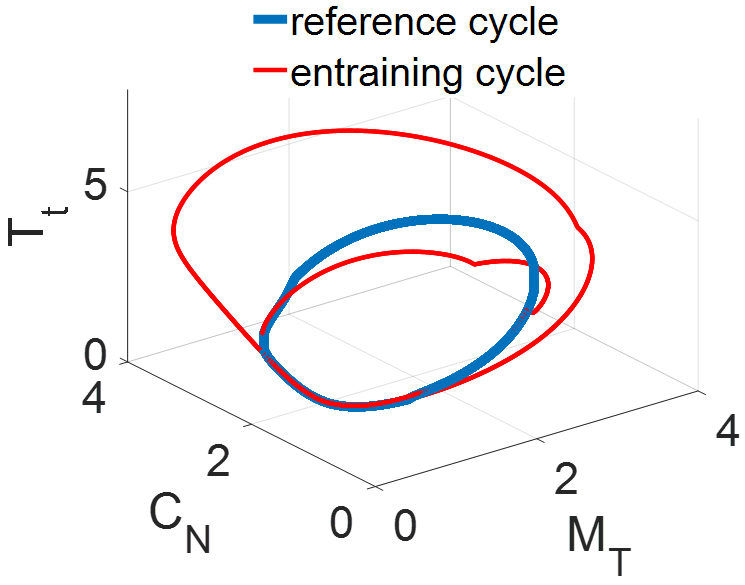}
			\caption{\small $M_{\rm T}$-$C_{\rm N}$-$T_t$ cycle}
			\label{fig:12_shift_Leloup_MT_CN_Tt}
		\end{subfigure}
		\begin{subfigure}[b]{0.20\textwidth}
			\includegraphics[width=\textwidth]{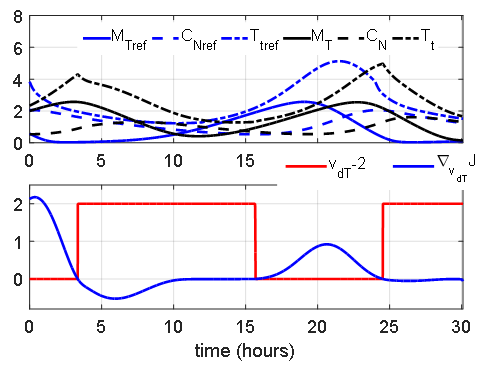}
			\caption{\small 16 hours shift}
			\label{fig:16_shift_Leloup2}
		\end{subfigure}
		\begin{subfigure}[b]{0.19\textwidth}
			\includegraphics[width=\textwidth]{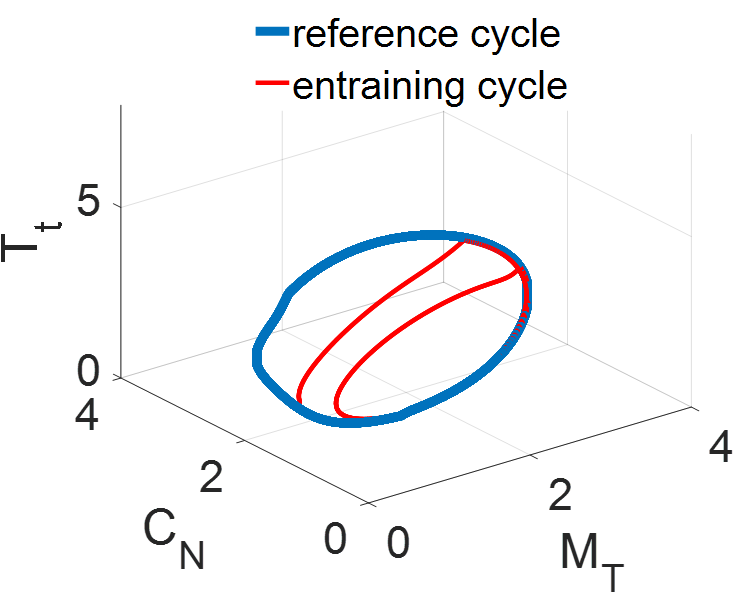}
			\caption{\small $M_{\rm T}$-$C_{\rm N}$-$T_t$ cycle}
			\label{fig:16_shift_Leloup_MT_CN_Tt}
		\end{subfigure}  
		\caption{\small Optimal entrainment cases of the Drosophila model.}
		\label{fig:Entrainment_cases_Leloup2}  
	\end{figure}

	Fig. \ref{fig:Leloup_Entrainment_initial} shows that $u_{\rm 2nd}$ is the best initial guess for GDA in the Drosophila model. The entrainment time costs in Fig. \ref{fig:Leloup_Entrainment_Time} show the maximum time cost of the optimal entrainment occurs at $\Delta_{\rm threshold}=$14 hours shift, with a value of 50 hours. In Fig. \ref{fig:Entrainment_cases_Leloup2}, we use the $C_{\rm N}$-$M_{\rm T}$-$T_{t}$ cycle to demonstrate the circadian oscillator of Drosophila. Similar with the mammalian and Neurospora model, the amplitude of circadian oscillator is enlarged when $\Delta_{\rm shift}<\Delta_{\rm threshold}$ and quenched when $\Delta_{\rm shift}>\Delta_{\rm threshold}$ during the minimum-time optimal entrainment. These phenomena imply that, in these models, the minimum-time optimal light delays and advances the circadian rhythm by enhancing and inhibiting the clock gene transcription and protein synthesis, respectively.
	
	Note that in the 8-hour shift case, the GDA converge to a light solution that is no longer a bang-off one with $\nabla_{v_{\rm dT}(t)}J=0$ in some intervals. We call this case as a {\it singular case}.
	
	\section{Discussions and Conclusions}
	
	
	In this paper, we solve the minimum-time entrainment problem in the high order mammalian, Neurospora, Drosophila circadian genes regulation model. We first obtain the optimal lighting strategies for the 1st- and 2nd-order reduced models. The gradient descent algorithm then searches for (locally) optimal solutions with those from the reduced models as initial guesses.
	
	\textbf{Impacts of initial guesses on gradient descent results:} The optimal light from the 2nd-order model works better on the GDA than other initial guesses in the Drosophila and Neurospora model. However, the results in the mammalian model show that the greedy advance strategy is the best initial guess for the GDA in a certain case. Therefore, in the solution procedure, we should initialize the GDA by both the greedy delay/advance strategy and the 2nd-order light strategy, and choose the one with the minimum time cost.
	
	\textbf{Impacts of the optimal light on circadian gene-protein:} Among all three models in this paper, the amplitude of the gene-protein oscillator is quenched or enlarged in every case. The results demonstrate that, during optimal entrainment, circadian delay and advance are closely linked with enlargement and quenching of the gene-protein oscillator in these models.
	
	Compared with the minimum-time entrainment, the delay/advance strategy takes longer time. However, the entrainment time of this strategy is less than $u_{\rm ref}$ and $u_{\rm 2nd}$ in mammalian and Neurospora models. This strategy is attractive as a feedback controller if the measurement of the circadian phase is available during the entrainment process. 

    \bibliographystyle{IEEEtran}
    \bibliography{Jiawei_circadian_ref,IEEEabrv}
	
\end{document}